\documentclass[10pt,journal,compsoc]{IEEEtran}
\ifCLASSOPTIONcompsoc
  \usepackage[nocompress,noadjust]{cite}
\else
  \usepackage{cite}
\fi
\ifCLASSINFOpdf
  \usepackage[pdftex]{graphicx}
\else
\fi
\usepackage{amsmath}
\usepackage{algpseudocode, algorithm}
\ifCLASSOPTIONcompsoc
  \usepackage[caption=false,font=footnotesize,labelfont=sf,textfont=sf]{subfig}
\else
  \usepackage[caption=false,font=footnotesize]{subfig}
\fi

\usepackage{stfloats}
\usepackage{url}

\usepackage{amsmath}
\usepackage{amssymb}
\usepackage{algpseudocode, algorithm}
\usepackage{booktabs}
\usepackage{footmisc}
\usepackage{hyperref}

\newtheorem{definition}{Definition}

\begin{document}
%
\title{Discriminatory Expressions to Improve Model Comprehensibility in Short Documents}
%
%
%
%

\author{Manuel~Francisco, Juan~L.~Castro%
\IEEEcompsocitemizethanks{\IEEEcompsocthanksitem M. Francisco and J. L. Castro are with the Department of Computer Science and Artificial Intelligence, University of Granada, 18071 Granada, Spain.}}

\markboth{Preprint sent to IEEE Computational Intelligence Magazine}%
{Francisco \MakeLowercase{\textit{et al.}}: Preprint sent to IEEE Computational Intelligence Magazine}
%



\IEEEtitleabstractindextext{%
\begin{abstract}
Social Networking Sites (SNS) are one of the most important ways of communication. In particular, microblogging sites are being used as analysis avenues due to their peculiarities (promptness, short texts...). There are countless researches that use SNS in novel manners, but machine learning has focused mainly in classification performance rather than interpretability and/or other goodness metrics. Thus, state-of-the-art models are black boxes that should not be used to solve problems that may have a social impact. When the problem requires transparency, it is necessary to build interpretable pipelines. Although the classifier may be interpretable, resulting models are too complex to be considered comprehensible, making it impossible for humans to understand the actual decisions. This paper presents a feature selection mechanism that is able to improve comprehensibility by using less but more meaningful features while achieving good performance in microblogging contexts where interpretability is mandatory. Moreover, we present a ranking method to evaluate features in terms of statistical relevance and bias. We conducted exhaustive tests with five different datasets in order to evaluate classification performance, generalisation capacity and complexity of the model. Results show that our proposal is better and the most stable one in terms of accuracy, generalisation and comprehensibility.
\end{abstract}

}

\maketitle

\IEEEdisplaynontitleabstractindextext

%
\IEEEpeerreviewmaketitle

\IEEEraisesectionheading{\section{Introduction}\label{sec:introduction}}

%
%
%
%
\IEEEPARstart{N}{owadays}, information flow is huge. From traditional media such as newspaper and television to blogs and forums, sources of information have grown to a point where we cannot conceive our reality without them. Social Media has become one of the most important ways to communicate. Users from all around the world can now manifest their opinions and share content regardless of their socio-economic status. This makes Social Networking Sites (SNS) a baseline to analyse people's interest and opinion trends.

Microblogging sites are prone to analysis applications such as environmental problem detection and sentiment analysis at different levels~\cite{perinan-pascual_detecting_2019,alharbi_twitter_2019}. They can be understood as a combination of blogging (since general public can access your history of publications) and instant messaging (because messages are supposed to be short enough to keep a fluid communication). Messages may include media such as images and videos.

Arguably, the most popular one is Twitter, with $126$ million daily active users as of 2018 and almost $6000$ tweets being sent per second~\cite{twitter_inc_q1_nodate, noauthor_twitter_nodate}. In recent years, Twitter has become a very important tool for research. The number of papers published regarding this topic has increased substantially since the platform was born in 2007 (see table~\ref{tab:twitterpapers}). However, \textit{tweets} show clear handicaps when performing analysis of their content that impoverish machine learning (ML) capacities: lack of context; abundance of misspelled words, contractions and acronyms; and new semantic units (like hashtags), among others. 

\begin{table}[tbp]
    \centering
    \caption{Number of publications that results from the search of the word ``Twitter'' in the title, abstract and/or keywords in Scopus. Updated February 9th 2021.}
    \label{tab:twitterpapers}
    \footnotesize
    \begin{tabular}{cccccccc}
        
        \toprule
        year & 2020 & 2019 & 2018 & 2017 & 2016 & 2015 & 2014\\
        \midrule
        no. papers & 4748 & 4627 & 4227 & 3754 & 3564 & 2934 & 2934\\
        \bottomrule
    \end{tabular}
    \newline\vspace*{2mm}\newline
    \begin{tabular}{cccccccc}
        \toprule
        year & 2013 & 2012 & 2011 & 2010 & 2009 & 2008 & 2007 \\
        \midrule
        no. papers & 2575 & 2304 & 1652 & 617 & n188 & 36 & 10 \\
        \bottomrule
    \end{tabular}
\end{table}

Along with these handicaps, another problem arises when considering domains where interpretability is required. Mainly, any decision that may have a social impact is subject to being audited and should be transparent and fair. For example, censor algorithms are a controversial topic regarding this matter, since preemptive closing of accounts limits free speech and should have explicit reasoning~\cite{phillips_moral_2018, hanff_alexander_cold_nodate}. Recommendation systems also present this kind of issue as filtering the content that is shown to a user may impact their interests and opinions, and even affect third parties~\cite{odair_beyond_2019,zhao_impact_2018,lee_impact_nodate,lee_when_2016}. For industries and businesses, complying with regulations and ethic constraints may also require interpretable models.

However in recent years, machine learning research has focused mostly in performance metrics (such as \textit{accuracy}, \textit{$f1$-score} or \textit{ROC AUC}), forgetting about interpretability. State-of-the-art language models (such as ELMo or BERT~\cite{peters_deep_2018,devlin_bert:_2018}), presented by AllenNLP and Google, and GPT-2 or GPT-3~\cite{radford_language_2019,brown_language_2020}, presented by OpenAI) are based on \textit{deep} techniques and thus they are not interpretable. These models are extremely powerful in terms of potential uses, and they have proven to be effective in tasks in which they were never trained for, such as language translation and code writing. To achieve this outstanding capabilities, they use techniques such as \textit{Transformers}, that are based on the concept of \textit{Attention}, but they require billions of hyperparameters that need to be optimised and fine-tuned to each specific task~\cite{vaswani_attention_2017}.

At the same time, researches are noticing the vast moral implications of taking high stakes decisions with ML models. Lately, numerous events and studies regarding interpretability are emerging, giving this field more relevance both between scientists and citizens, as can be seen in the list gathered by \cite{carvalho_machine_2019}. Table~\ref{tab:interpretabilitypapers} reflects the increasing number of papers related to interpretability.

\begin{table}[tbp]
    \centering
    \caption{Number of publications that results from the search of the word ``interpretability'' in the title, abstract and/or keywords in Scopus. Updated February 9th 2021.}
    \label{tab:interpretabilitypapers}
    \begin{tabular}{cccccccc}
        
        \toprule
        year & 2020 & 2019 & 2018 & 2017 & 2016 & 2015 & 2014 \\
        \midrule
        no. papers & 2181 & 1641 & 1037 & 809 & 615 & 570 & 477 \\
        \bottomrule
    \end{tabular}
\end{table}

Instead of developing new interpretable models, there is a growing tendency to \textit{explain} current black-box models with subordinated ones (\textit{post-hoc} models trained to give reasons on why the original model behaves that way); this approach may preserve bad practices over time that can lead to potential harm~\cite{rudin_please_2018}.

There are several ways to build ML models, but the typical procedure includes: several preprocessing steps (obtaining features, removing outliers, reducing dimensionality...), training the model and validating it. It is common belief that model accuracy is directly related to its complexity. \cite{rudin_please_2018} established that this statement is not necessarily true, since well-structured significant features often yields the same results regardless of the complexity of the classifier. In this paper, we will focus on the features used to train a model that is required to be interpretable, to obtain a feature selection method that is able to:

\begin{enumerate}
    \item Improve classification performance using a low number of features.
    \item Improve generalisation capacity using a low number of features.
    \item Improve comprehensibility by using less but more meaningful features.
\end{enumerate}

We will measure these items in terms of mean value and stability through different classifiers and datasets to ensure that the method is versatile enough. We refer to generalisation capacity as the ability of a model to obtain good classification performance when training with a number of instances much lower than the number of validation samples. This is the typical situation when working with social media.

We believe that we can use expressions biased towards a class in order to achieve our targets. An expression will be a sequence of words (not necessarily contiguous) with a specific order. This is partially what linguists consider when analysing text: not only the words but also the order in which they are presented. In fact, they also take into account stop words, on the contrary to machine learning models where stop words are removed in most scenarios while preprocessing. \cite{moreo_high-performance_2012} tried a similar approach with success, although in a different domain. Additionally, we will need a new measure that is able to maximise statistical relevance and discriminatory performance in order to rank the expressions in a massive search space.

In this article, we present an algorithm that is able to obtain and select expressions from short texts like tweets. Since they will be sequence of natural language words, features will be easy to interpret. Moreover, selected expressions will be biased towards a class to ensure classification performance while decreasing model complexity. Thus, we introduce a ranking method called \textit{CF-ICF} that is able to rank words taking into consideration the bias towards the class whose expressions we are looking for, while maximising statistical relevance.

Our results show that our proposal has one of the best classification performance between the compared methods. They also show that it has one of the lowest deviations and it manages to reduce the complexity of the resulting models in some cases, hereby improving comprehensibility. Decreasing model complexity is important because not all models are interpretable in practice (e.g. a huge decision tree with a lot of leaves is potentially interpretable, but it is impossible to do it effectively).

\vspace{1cm}
The rest of this paper is organised as follows. In section~\ref{sec:sota} we review literature related to feature selection methods and introduce concepts on interpretability. Section~\ref{sec:preliminary} shows results on a preliminary study to check if current interpretable models are comprehensible enough. In section~\ref{sec:dealgorithm} we present our algorithm of Discriminatory Expressions, as well as a ranking method based on the classic \textit{TF-IDF}. We performed several experiments as mentioned in section~\ref{sec:experiments} and we discuss results in section~\ref{sec:results}. Section~\ref{sec:limitations} present our main contributions and limitations of our work. Sections~\ref{sec:conclusions} and~\ref{sec:futurework} sum up our proposal and present several lines of future work.
\section{Related Work}
\label{sec:sota}
Text mining research in Social Media has become very relevant in the last few years, as they are ground for quite a few computer science applications. From analysing human behaviour to stock prediction, there are a lot of ongoing projects and researches based on topic detection~\cite{perinan-pascual_detecting_2019, castellanos_formal_2017, nassar_enhancing_2016, xie_topicsketch:_nodate}, measuring user's influence~\cite{alsaig_context_2019, qasem_detection_2016, jorgens_exploring_2016, subbian_mining_2016}, sentiment analysis~\cite{alharbi_twitter_2019,xiaomei_microblog_2018, rosenthal_semeval-2017_2017, saif_contextual_2016, supriya_twitter_2016}, opinion mining~\cite{li_mining_2016,gasco_beyond_2019, takahashi, kumar_analysis_2016}, and text summarisation~\cite{qian_social_2019,arroyo-fernandez_language_2019}, among others. However, NLP processing techniques may not be good enough under \textit{microblogging} circumstances, since they result in highly-dimensional and very sparse feature matrices~\cite{zheng_learning-based_2018}. Applications like authorship identification (duplicate accounts, inter-network identification...) have demonstrated that traditional methods are not feasible in short-message contexts like Twitter, as they tend to assume a minimum text extension under which models would not be suitable enough. Alternative techniques like \textit{style markers} have been proved better~\cite{green_comparing_2013}. Other document-pivot methods present similar problems, and recent approaches based on co-occurrence, \mbox{\textit{TF-IDF}} and/or pattern recognition techniques (such as FP-Growth~\cite{han_mining_2000}) are being used for this purpose~\cite{li_twevent:_2012, sayyadi_graph_2013}.

In any case, when dealing with natural language models, the feature space is usually extremely wide. Normally, a set of documents has a few thousand unique words that are used as features, despite many of them can be considered noisy or not relevant at all. It is required to select and/or recombine them, not only to decrease the complexity of the problem but also to improve classification performance~\cite{wang_supervised_2019}.

There are several classes of feature selection (FS) methods, but they are generally grouped into three categories~\cite{xue_particle_2013}:
\begin{itemize}
    \item Filtering methods. Given a set of features, they apply an evaluation function and select the $k$ best, where $k$ is an hyperparameter. They score each feature taking into consideration different aspects of them, like document frequency (DF) or TF-IDF.
    \item Wrapper methods. Given a set of features, they select different subsets of them to train a classifier and check out how good its performance is. Since selected features are tested directly within the classifier, they normally achieve better performance than filters.
    \item Hybrid methods. They combine both techniques: first, they perform a filter to reduce the number of features; then, optimal subset is computed by feeding a classifier.
\end{itemize}

There is another group of feature selection techniques called \textit{embedded methods}~\cite{liang_text_2017}. These methods are inherent to the classification stage, since they take part in the training process (e.g. decision trees), and they are usually not considered independently.

There are loads of filtering methods available and backed by the scientific community. We have selected a few of them based on their current relevance, goodness metrics and community acceptance. They are presented below:


\paragraph{CHI2 (chi-square)} $\chi^2$ is one of the most popular filtering methods for feature selection problems. It is possible to use the statistical test in order to check the independence of two events ($p(AB) = p(A)p(B)$), in this case, a feature and a class.
\begin{equation}
\label{eq:chi2}
    \chi^2_{(t,c)} = \frac{D\times\left[p(t|c)p(\overline t| \overline c)-p(\overline t|c)p(t|\overline c)\right]^2}{p(t)p(\overline t)-p(c)p(\overline c)}
\end{equation}
$\chi^2$ is defined for text classification through equation~\ref{eq:chi2}, where: $D$ stands for the total number of documents, $t$ is a feature and $c$ is a class~\cite{rutkowski_artificial_2008,senthil_kumar_b_different_nodate,zheng_feature_2004}.


\paragraph{Information Gain (IG)} IG is based on entropy (information theory). It measures the gain of a feature with respect to a given class (decrease in entropy between considering the feature or not)~\cite{forman_extensive_2003}. IG is defined as stated in equation~\ref{eq:ig}, where $t$ is a feature and $c$ a class~\cite{caropreso_learner-independent_nodate,ding_hybrid_2018,deng_feature_2019,largeron_entropy_2011}.

\begin{equation}
    \label{eq:ig}
    IG_{(t, c)} = p(t|c)\log\frac{p(t|c)}{p(t) p(c)}+p(\overline t|c)\log\frac{p(\overline t|c)}{p(\overline t) p(c)}
\end{equation}


\paragraph{Mutual Information (MI)} MI measures how much information two variables share, in this case, a feature $t$ and a class $c$. It is defined in equation~\ref{eq:mi}~\cite{deng_feature_2019,al-salemi_rfboost:_2016}. It can be proved equal to IG for binary problems~\cite{senthil_kumar_b_different_nodate}.

\begin{equation}
    \label{eq:mi}
    MI_{(t,c)} = \log\frac{p(t|c)}{p(t) p(c)}
\end{equation}
with $t$ and $c$ being a feature and a class, respectively.


\paragraph{Odds Ratio (OR)} OR measures the probability of a term $t$ and a class $c$ co-occurring normalised by the probability of $t$ occurring in other classes~\cite{al-salemi_rfboost:_2016,zheng_feature_2004}.

\begin{equation}
    OR_{(t,c)}=\log\frac{p(t|c)(1-p(t|\overline c))}{(1-p(t|c)) p(t|\overline c)}
\end{equation}
where $t$ is a feature and $c$ is a class.


\paragraph{Expected Cross Entropy (ECE)} ECE computes the distance between the class $c$ distribution and the class distribution co-occurring with the feature $t$. \cite{wu_improved_2015} defined ECE as in equation~\ref{eq:ece}.

\begin{equation}
    \label{eq:ece}
    ECE_{(t,c)} = p(t)\left(p(c|t)\log\frac{p(c|t)}{p(c)}+p(\overline c|t)\log\frac{p(\overline c|t)}{p(\overline c)}\right)
\end{equation}


\paragraph{ANOVA F-value} It is used to check if there is a significant difference between the variance of two variables~\cite{misangyi_adequacy_2016}. In one-way ANOVA, the $F$-statistic is defined as in equation~\ref{eq:anova}:
\begin{equation}
    \label{eq:anova}
    F\text{-statistic} = \frac{\text{variance between groups}}{\text{variance within groups}}
\end{equation}

    
\paragraph{Galavotti-Sebastiani-Simi coefficient (GSS)} \cite{galavotti_experiments_2000} proposed a simplified version of $\chi^2$ given by equation~\ref{eq:gss}~\cite{largeron_entropy_2011}.
\begin{equation}
    \label{eq:gss}
    GSS_{(t,c)}=p(t|c)p(\overline t|\overline c)-p(t|\overline c)p(\overline t|c)
\end{equation}

where $t$ is a feature and $c$ is a class. 

All these filters present similar drawbacks. Mainly, features are selected regardless of the classifier, so the selected feature subset may not be ideal for every classification stage. However, they are quicker computing the aforementioned subset than other categories of FS mechanisms.

On the other hand, wrappers assess the relevance of each subset of features within the context of a given classifier. Goodness metrics for the classification model determine how good each feature subset is. Hence, the FS method will choose the subset of features that yields the best performance. There are tons of generic wrappers for feature selection, such as~\cite{dy_feature_2004,shah_data_2004,meiri_using_2006,hans_shotgun_2005}. They usually achieve better performance than filtering methods.

However, the complexity of evaluating each subset is quite high (exponential). To overcome this disadvantage, they are usually combined with metaheuristics such as genetic algorithms, ant colony optimisation, particle swarm optimisation and/or iterated local search~\cite{ak_feature_1997}. Despite the complexity of performing a search, they still are suboptimal approaches and classifier-dependant.

Wrapper approaches can also be combined with filters in order to narrow the space search to a promising area (hybrid methods). Hence, they are suitable for early convergence (which can lead to local extrema) and prone to overfitting (at least with small datasets)~\cite{lee_competitive_2019}.

Since 2012, \textit{state-of-the-art} is abandoning these \textit{handcrafted} methods in favour of auto-encoded features~\cite{minaee_deep_2020}. Deep learning techniques automate the pipeline building process by learning features and subsequent classification rules at the same time. Nowadays, their popularity is outstanding. \cite{minaee_deep_2020} lists several categories of deep models, including the following:
\begin{itemize}
    \item Recurrent Neural Networks (RNNs) and their famous variant, Long-Short Term Memory (LSTM). They are designed to extract dependencies and patterns over time within sequence of words.
    \item Convolutional Neural Networks (CNNs) try to capture invariant structures over documents, such as expressions or figures of speech.
    \item Attention, which focuses on correlation between words by weighting each word with respect to others in the document.
    \item Transformers, built upon the \textit{Attention} concept to overcome the sequential limitation of RNNs.
\end{itemize}

These techniques yielded language models such as ELMo~\cite{peters_deep_2018}, BERT~\cite{devlin_bert:_2018}), GPT-2~\cite{radford_language_2019} and GPT-3~\cite{brown_language_2020}. In its most general form, the mechanism relies on representing words (and even context) as multi-dimensional vectors (\textit{embeddings}) in a manner in which those related to similar words (in terms of meaning and/or related context) are closer in the space. This allows that certain operations can be made over them with acceptable accuracy, such as adding restrictions or substracting partial meanings (arguably, the most popular example is the one were $\overrightarrow{\text{king}} - \overrightarrow{\text{man}} + \overrightarrow{\text{woman}} = \overrightarrow{\text{queen}}$).

After computing such \textit{embeddings}, they reduce dimensionality, by \textit{encoding} hidden patterns within the weights of intermediate layers, until (1) they can be handled by an arbitrary classifier or until (2) dimensionality is reduced to the classes themselves.

We cannot consider these approaches since they behave as black-box models and cannot be interpreted. In fact, most of the techniques we have reviewed regarding feature selection were focused in reducing dimensionality and improving classification performance. However, little attention has been paid to \emph{how interpretable} are the selected features.

It is possible to obtain a set of features that are easier to understand by humans without significantly affecting classification performance~\cite{rudin_please_2018}. In order to ensure model integrity, they should be transparent~\cite{fatml_principles_nodate}. Furthermore, for some text classification problems, there are features that should not be selected at all, at least in \textit{microblogging} context. For examples, those related to typing errors or contractions, when these are due to the limit of characters.

There is no general consensus on how to measure interpretability. This concept, along with similar ones like \textit{comprehensibility}, \textit{understandability} and so on, are usually synonyms~\cite{alonso_looking_2009}. However, there is kind of a distinction between those related to the ability to read the model (\textit{interpretability}) and those related to the human capacity to understand it (\textit{comprehensibility})~\cite{mencar_interpretability_2008}. \cite{doshi-velez_towards_2017} defined \textit{interpretability} as ``\textit{the ability to explain or to present in understandable
terms to a human}''. They elaborated a taxonomy on how to evaluate interpretability:
\begin{enumerate}
    \item Application level, consisting in an expert evaluation of the model itself.
    \item Human metrics, where any human can perform such evaluation without the need of being a domain expert. It is normally performed by comparison with other models/explanations.
    \item Proxy tasks, when we make the assumption that the user understands the model and we only compare parameters within it (e.g. depth of a decision tree).
\end{enumerate}

It is particularly interesting to perform evaluations at application level. Nevertheless, this is a very expensive task that most studies do not contemplate unless it is strictly necessary.

Our proposal focuses in a functional evaluation (proxy task). We based it in \cite{moreo_high-performance_2012}, who tried a similar approach in a \textit{frequent answered questions (FAQ)} retrieval method and continued with~\cite{moreo_towards_2017, moreo_learning_2013, moreo_faqtory:_2012}. However, this model computes the whole set of \textit{minimal differentiator expressions}, which is computationally eager and it is only viable in closed domains like the one proposed by the authors. It is not a valid solution for an open, always-growing environment like Social Media.
\section{Preliminary Study}
\label{sec:preliminary}

There are interpretable classifiers, such as decision trees (DT) or $k$-nearest neighbours (kNN), that can be used to build interpretable models. However, this is only doable in theory as actual models are too complex to be understandable. We justify in this section the necessity of new approaches that yield more comprehensible models.

There is a neat question regarding how many aspects can be handled by humans. In 1956, \cite{miller_magical_1956} established precedent in what would be considered \textit{working memory}. The article shows that $7\pm2$ is the number of chunks that a person can remember for a short time. The author referred to that number as a unitary measure but not in terms of minimum possible unit. That means that any human can handle between $5$ and $9$ concepts, situations, facts, melodies, etc., rather than words, movements or sounds (minimum expression). Further research proved that this number could be smaller~\cite{cowan_magical_2001}.

Interpretability is not a general concept, and it should be understood within the domain of the studied problem~\cite{rudin_please_2018,freitas_comprehensible_2014, huysmans_empirical_2011}. Consequently, there is no formula that can be used to measure how interpretable a model is. In order to evaluate it, we need to consider its characteristics.

We selected three classifiers that can be considered interpretable~\cite{molnar_interpretable_2018}. They are widely used and capable of handling different type of relations between features:
\begin{enumerate}
    \item $k$-nearest neighbours (kNN). Given an instance, it retrieves the $k$ most similar instances within the training set. The resulting class (output) would be the mode between the classes of the $k$ closest instances. It can be considered self-explanatory, as it only compares through individuals of the same type. The complexity can be measured in two dimensions, being (1) the number of neighbours ($k$) and (2) the individual itself (number of features required to represent the document).
    \item Decision Tree (DT). The particulars regarding how a DT works depend on the algorithm used to build it. In general, they split data into several partitions regarding some criteria. In order to interpret them, one will start in the root node. Each node represent a condition over an attribute with two (or more) possible outcomes (edges). Leaf nodes stand for the output of the classification process. Each path from the root to the leaves can be represented as a rule of \textit{AND clauses}, each of them being a node. They can be measured in two dimensions: (1) number of leaves (equivalent to \textit{number of rules}) and (2) length of the path (equivalent to \textit{length of the rule}).
    \item Random Forest (RF). They are collections of Decision Trees. Each DT will be trained with a subset of the training instances. The output can be obtained through the mode or the weighted addition of the individual outputs from decision trees. It is possible to interpret this classifier as individual decision trees with different perspectives of the same reality, as it would happen with a jury of humans. The complexity of the model can be measured in three dimensinos: (1) number of trees, (2) number of leaves of each tree and (3) length of each path.
\end{enumerate}

There are another two dimension that we need to consider to measure complexity. They affect all the classifiers mentioned above: comprehensibility and number of features. It is easy to argue that, within the context of our problem, features are as understandable as they can be, since they are sets/sequences of words. That leaves us with number of features.

We have conducted several experiments to measure the complexity of the models built with the aforementioned classifiers. We followed a $5$-cross fold validation scheme using python 3.8.1 and other libraries (see section~\ref{sec:experiments}). All tweets were tokenised and stemmed before selecting features.

\begin{table}[tbp]
    \centering
    \caption{Number of features required to achieve the median $f1$-score (approx. $0.54$) for each feature selection method. Results shown are arithmetic mean and standard deviation through different folds (iterations), classifiers and datasets. See section~\ref{sec:experiments} for more details.}
    \label{tab:preliminary_nf}
    \begin{tabular}{lrrr}
    \toprule
    fs method & mean $f1$-score & std $f1$-score & number of features \\
    \midrule
    CHI2   & 0.551399  & 0.111377 & 20     \\
    ECE    & 0.557570  & 0.061533 & 100    \\
    ANOVA  & 0.564128  & 0.074548 & 20     \\
    IG     & 0.546856  & 0.108496 & 20     \\
    MI     & 0.550547  & 0.107467 & 20     \\
    OR     & 0.553999  & 0.071509 & 100    \\
    \bottomrule
    \end{tabular}
\end{table}

\begin{table}[tbp]
    \centering
    \caption{Out-of-the-box mean performance and complexity measure for several classifiers. Results are obtained as a mean through different feature selection methods and datasets. See section~\ref{sec:experiments} for more details.}
    \label{tab:preliminar}
    \begin{tabular}{lrl}
        \toprule
        Classifier & $f1$-score & Complexity Measure \\
        \midrule
        kNN & 0.5481 & 20 features             \\
        DT  & 0.6277 & 366 rules of 12 clauses  \\
        RF  & 0.6501 & 100 trees                \\
        \bottomrule
    \end{tabular}
\end{table}

Results show that there is still room for improvement. Table~\ref{tab:preliminary_nf} shows that it is necessary to deal with at least 20 features in order to have acceptable performance metrics (above median value). Table~\ref{tab:preliminar} shows that default methodology yields models that are too complex to be actually comprehensible, despite that they are interpretable models. Since human capacities are limited, it is necessary to find new methods that enhance complexity measures.

\section{Discriminatory Expressions}
\label{sec:dealgorithm}
Taking into account aforementioned reasons, we present in this section our proposal to obtain significant feature sets that improve model comprehensibility in microblogging contexts.
Words are a tool used to abstract reality. As such, we are the ones that give sense to them, even creating new meanings. However, they require context: it is impossible to communicate effectively without sentences. Given a word, depending on the words that precede or follow it, the meaning can vary from one end to the other. Meaning is not only influenced by the order of the words but also by adjectives, prepositions and even \textit{stop words} (that are usually removed in preprocessing steps). Some expressions are prone to be interpreted in several ways, but they maintain the bias to some extent. We rely on this bias to find the set of expressions that defines a class.

Let $X = \{d_1, d_2, ..., d_n\}$ be a set of documents where each document $d \in X$ belongs to a class $C$. We consider $S$ as the set of \textit{stop words}.

\begin{definition}[Expression]
Given a document $d \in X$ as a sequence of words $d = (t_1, t_2, ..., t_n)$, $e$ is said to be an expression of the document, noted as $e \textit{ expr } x$, iff:
\begin{enumerate}
    \item It is not composed strictly by \textit{stop words}.
    \item All words of the expression can be found in the document and the order is preserved ($e$ is a regular expression that matches the document).
\end{enumerate}

\begin{equation}
    e \textit{ expr } d \longleftrightarrow \left\{
        \begin{array}{l}
            \exists t_i \in e : t_i \not\in S\\
            \wedge \\
            /t_1*t_2*...*t_n/ \text{ matches } d
        \end{array}
    \right.
\end{equation}
\end{definition}

\begin{definition}[Discriminatory expression] Given $r$ (minimum relevance or recall) and $p$ (minimum precision), an expression $e$ is said to be \mbox{$(r,p)$-discriminatory} for the class $C$, noted as $e \textit{ dexpr}_{r,p}\text{ } C$ iff:

\begin{enumerate}
    \item Recall of $e$ for the class $C$ is over a given threshold $r$.
    \item Precision of $e$ for the class $C$ is at least $p$.
    
\begin{equation}
    e \textit{ dexpr}_{r,p}\text{ } C  \longleftrightarrow \left\{
        \begin{array}{l}
            \frac{|e \textit{ expr } d_i : d_i \in C|}{|C|} > r \\
            \wedge \\
            \frac{|e \textit{ expr } d_i : d_i \in C|}{|e \textit{ expr } d'_i : d'_i \in X|} > p
        \end{array}
    \right.
\end{equation}
\end{enumerate}

\end{definition}

Despite expressions may be an intuitive way to look at \textit{microblogging} documents, the search space is too complex to be explored completely. There are several ways to obtain suboptimal solutions to this problem (metaheuristic techniques). We used a greedy approach with a custom ranking function based on TF-IDF.

\subsection{Ranking Method: \textit{CF-ICF}}
\label{sec:cficf}
We build discriminatory expressions from the words that are present in the documents of a class. Usually, there are several expressions that can accomplish our criteria, but a few of them are more useful than the rest. As we want to maximise statistical relevance and discriminatory performance, we propose a ranking method based on \textit{TF-IDF} that takes into consideration the class whose expression we are looking for.

Let $X$ be a set of documents $X = \{d_0, d_1,..., d_n\}$ where $n = |X|$. $L$ is a binary property that can be present (or not) in each document, such that for a given document $d_i \in X$, $L(d_i) \in \{0,1\}$. We will now define the sets $X_L^+ \subseteq X$ and $X_L^- \subseteq X$ such that:

\begin{align}
    X_L^- = \{d \in X : L(d) = 0\} \\
    X_L^+ = \{d \in X : L(d) = 1\}
\end{align}
where $n_L^+$ and $n_L^-$ stand for $|X_L^+|$ and $|X_L^-|$, respectively.

The function $f(t, d)$ yields the number of times that a word $t$ appears in the document $d$. From here, we can define classic \textit{TF-IDF} as follows:

\begin{align}
    tf(t, d, X) &= \frac{f(t, d)}{\max\{f(t, d'), \forall d' \in X\}} \\
    df(t, X) &= |\{d \in X : t \text{ in } d\}| \\
    idf(t, X) &= \log\frac{n}{df(t,X)} \\
    tfidf(t,d,X) &= tf(t,d,X)idf(t,X)
\end{align}

Now, let $cf(t, L)$ be a function that returns the number of times that the word $t$ appears in the documents of $X_L^+$ and let $d_L = \bigcup_{d \in X_L^+}d$ (meaning that $d_L$ is the result of concatenating all the documents in $X_L^+$). We can define $cf$ as a function of $tf$ using the concatenated $d_L$ as follows:

\begin{equation}
    cf(t, L) = \sum_{d \in X_L^+} f(t, d) = tf(t, d_L)
\end{equation}

In the same way, let $n_L^-(t) = |\{d \in X_L^- : t \text{ in } d\}|$ and $n_L^+(t) = |\{d \in X_L^+ : t \text{ in } d\}|$. We can also express \textit{IDF} as follows:

\begin{equation}
    idf(t, X) = \log\frac{n}{n_L^-(t) + n_L^+(t)} \\
\end{equation}

Now, $idf$ can be modified to define $icf$, as shown below:

\begin{align}
    icf(t, L, X) &= \log\frac{n}{n_L^-(t) + 1} \\
    \label{eq:cficf}
    cficf(t, L, X) &= cf(t, L)icf(t,L,X)
\end{align}

The proposed ranking method can be seen in equation~\ref{eq:cficf}. It is straightforward to notice that

\begin{align}
    idf(t, X) &= icf(t, L, x) \Leftrightarrow n_L^+ = 1 \\
    cficf(t, L, X) &= \sum_{d \in X'}tfidf(t,d,X')
\end{align}

where $X' = \{d_L\} \cup X_L^-$. We can now study the relation between both methods:
\begin{enumerate}
    \item If $t$ is only in one document of $X_L^+$ (or none), then: 
    \begin{equation}
        cficf(t,L,X) = \sum_{d \in X_L^+}tfidf(t,d,X)
    \end{equation}
    
    \item If $t$ is in more than one document of $X_L^+$, then:
    \begin{equation}
        cficf(t,L,X) > \sum_{d \in X_L^+}tfidf(t,d,X)
    \end{equation}
    
    Note that if $t$ is in several documents of $X_L^+$, then \textit{CF-ICF} only depends on the number of times that this word appears in those documents. If $t$ only appears in $X_L^+$, \textit{ICF} will be the maximum possible ($\log n$), which implies that the maximum possible value for \textit{CF-ICF} will also be $\log n$, since \textit{cf} is normalised between $[0,1]$ and it will only be achieved if exists a word $t$ that is the most frequent in the class $L$ and it is not present in the documents of $X_L^-$.
\end{enumerate}

\begin{figure}
    \centering
    \includegraphics[width=\linewidth]{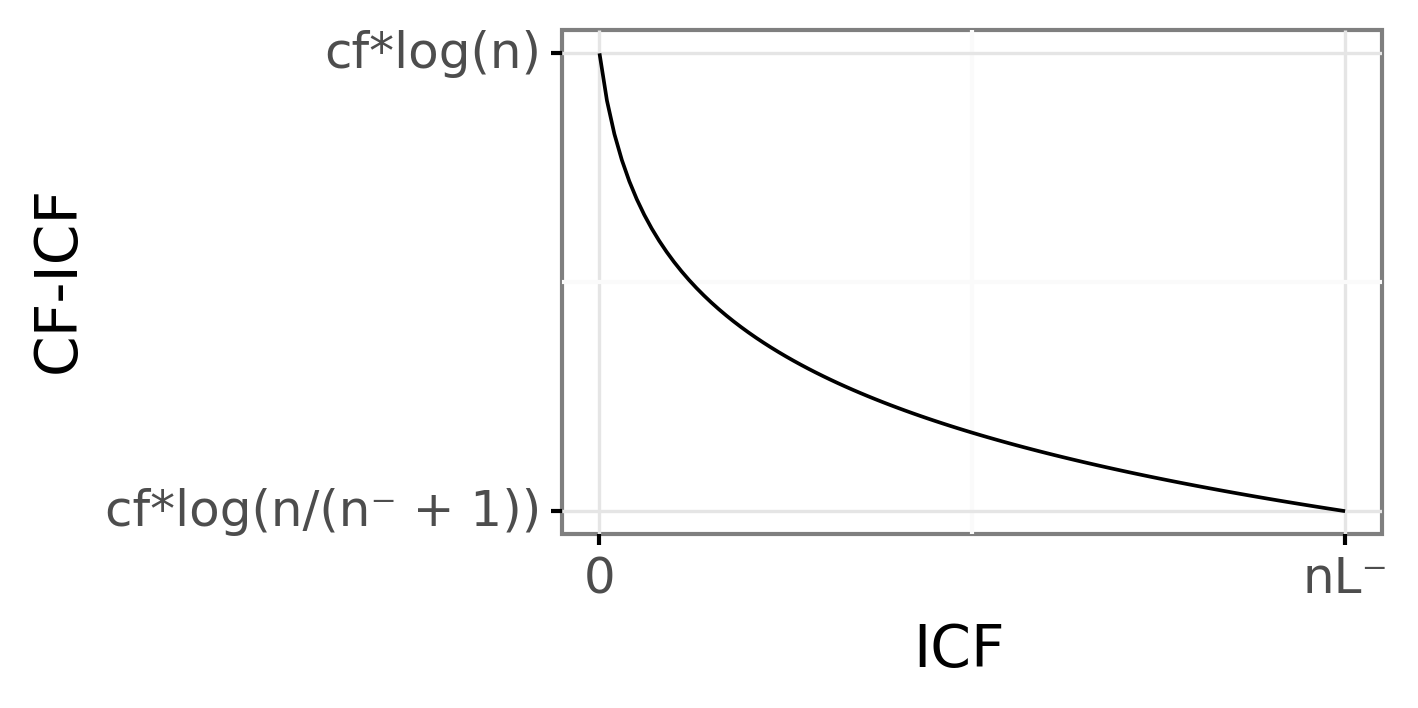}
    \caption{Behaviour of the \textit{CF-ICF} ranking method when keeping $cf$ fixed. It can be seen that the function decrease faster than a linear one.}
    \label{fig:fixedCF}

    \includegraphics[width=\linewidth]{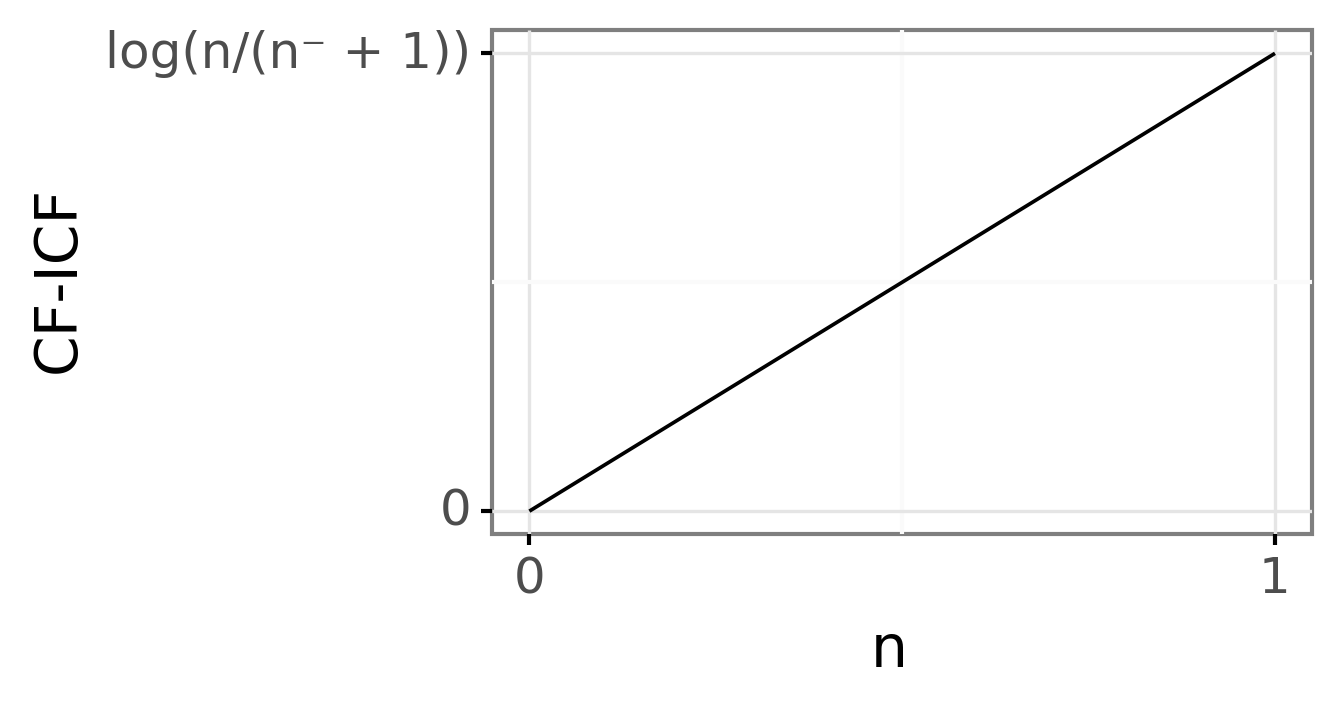}
    \caption{Behaviour of the \textit{CF-ICF} ranking method when keeping $icf$ still, showing that function grows linearly.}
    \label{fig:fixedICF}
\end{figure}

Figures~\ref{fig:fixedCF} and \ref{fig:fixedICF} describe the behaviour of \textit{CF-ICF} when keeping $cf$ and $icf$ fixed, respectively. It shows that is more important for the expression to be discriminatory than relevant ($icf$ decreases faster than $cf$ grows).

\subsection{Discriminatory Expressions Algorithm}
The algorithm~(\ref{dealgorithm}) we propose for feature extraction will compute a set of discriminatory expressions $D$ taking into consideration that each expression needs to meet some frequency and distinguishability criteria with respect to a class. In other words, each expression $e$ should be skewed towards a class, such that the frequency in which that expression appears in the class exceeds a certain threshold ($r$) meanwhile the ratio between the matches in the class and the total number of matches is above a given boundary ($p$).

\begin{figure}
    \begin{algorithmic}[1]
        \Require Training set of documents $X$, vector of labels $y$ for each document, minimum required precision $p$, minimum required recall $r$, maximum number of words in an expression $\alpha$, set of stop words $\textit{STOPW}$
        \Ensure Set of discriminatory expressions $D$
        
        \State $D \gets \emptyset$ \Comment{Initialise set of DE}
        
        \ForAll{$d \in X$}
            \State $o \gets $ \Call{queue}{} \Comment{Initialise queue of candidates}
            \State $t \gets$ \textbf{tokenize} $d$ 
            \State \textbf{stem} words from $t$
            \State \textbf{sort} elements in $t$ by \textit{cficf}
            \State \textbf{put} all elements of $t$ in $o$
        
            \Statex
            \Repeat
                \State $e \gets$ \textbf{pop} $o$
                
                \State $tp \gets $ \textbf{count} $e$ matches in $\{X_i : y_i = \textit{True}\}$
                \State $fp \gets $ \textbf{count} $e$ matches in $\{X_i : y_i = \textit{False}\}$
                \Statex
                \State $R \gets \frac{tp}{|\{y_i : y_i = True\}|}$
                \Statex
                \State $P \gets \frac{tp}{tp + fp}$
                \Statex
                \Statex
                \If{$R \geq r$}\Comment{Check if e is relevant enough}
                    \If{$P \geq p$ and $e \not\subseteq\textit{STOPW}$} \Comment{Check if e is discriminatory enough and discard those that are entirely stop words}
                        \State \textbf{accept} $e$
                        \State $D \gets D \cup e$
                    \ElsIf{$|e| < \alpha$} \Comment{Keep exploring until maximum length}
                        \State n $\gets$ $e \times t$
                        \State \textbf{put} all elements of $n$ in $o$
                    \EndIf
                \EndIf
            \Until{$e$ is accepted}
        \EndFor
        \Statex
        \State \Return $D$
    \end{algorithmic}
    \caption{Discriminatory Expressions algorithm}
    \label{dealgorithm}
\end{figure}

Once we have the set of discriminatory features, we can transform each document to a binary vector $(v_i)$ where each component $v_i$ stands for the appearance of the $i$-th discriminatory expression in the document.

Our python implementation can be found in our project's repository\footnote{\url{https://github.com/nutcrackerugr/discriminatory-expressions}}.
\section{Experiments}
\label{sec:experiments}
We conducted experiments to evaluate the performance of our proposal in terms of classification performance, generalisation capacity and comprehensibility. Through this section, we introduce the datasets we used to run our experiments, along with the specific settings of these.

\subsection{Datasets}
We used several popular datasets to test our proposal within different contexts and topics. Since it is not feasible to generate synthetic datasets (we extract and select the relevant features directly from plain text), we chose them taking into account dataset diversity (airlines sentiment, IMDB reviews, gender classification...), preciseness and acceptance within scientific community. 

All of these datasets are publicly available, except for~\cite{villena-roman_tass_2013} which can be requested to the authors through their website. Although they all have different feature set, we only used plain text and target class in our experiments.
\begin{itemize}
    \item US Airlines Sentiment. A collection of approximately $15$k tweets from \textit{Crowdflower's Data for Everyone library}\footnote{\url{https://www.figure-eight.com/data-for-everyone}\label{f8}} tagged for sentiment towards US airlines. It contains three classes (positive, neutral and negative, with $2363$, $3099$ and $9178$ instances, respectively) and a total of $14640$ tweets.
    \item Twitter User Gender. A dataset of around $25$k tweets labelled as \textit{male}, \textit{female}, \textit{brand} or \textit{unknown} from \textit{Crowdflower's Data for Everyone library}\footref{f8}. Distribution of instances between classes is as follows: $6700$ females, $6194$ males, $5942$ brands and $1117$ unknown samples for a total of $19953$ tweets.
    \item Sentiment140 dataset~\cite{go_twitter_2009}\footnote{\url{https://www.kaggle.com/kazanova/sentiment140}}. $1.6$M instances tagged as \textit{positive} or \textit{negative} sentiment, with $800$k instances for each class (balanced). 
    \item Sentiment Labelled Senteces (SLS) dataset (IMDB subset)~\cite{kotzias_group_2015}\footnote{\url{https://archive.ics.uci.edu/ml/datasets/Sentiment+Labelled+Sentences}}. A collection of 3000 instances of sentences labelled into positive and negative classes (balanced).
    \item TASS Sentiment Analysis dataset~\cite{villena-roman_tass_2013}\footnote{\url{http://tass.sepln.org/2017}}. A collection of $7$k Spanish tweets gathered by \textit{SEPLN}, classified in positive, neutral and negative sentiment. Each class has a total of $2884$, $2153$ and $2182$ tweets, respectively.
\end{itemize}

\subsection{Feature Selection Methods and Classifiers}
We compare our algorithm with several filter methods revised in Related Work (see section~\ref{sec:sota}). These are $\chi^2$ (CHI2), Information Gain (IG), Mutual Information (MI), Odds Ratio (OR), Expected Cross Entropy (ECE), ANOVA F-value (f-ANOVA) and \textit{Galavotti-Sebastiani-Simi} (GSS) coefficient. We described them in section~\ref{sec:sota}.

We used four classifiers that are widely considered interpretable. $k$-nearest neighbours (kNN), Decision Trees (DT), Random Forest (RF) and Logistic Regression (LR). We did not perform hyperparameter tuning on any of them. Although many authors consider LR an interpretable technique, we only used it to check accuracy within a linear model. It is possible to interpret how LR works, but we think it will not be easy for non-technical users to comprehend how it weights the features.

On the contrary, we chose RF since (1) it is straightforward to understand it once you know how DT works, (2) it is easier to comprehend by non-expert users with the analogy that each DT is an individual with its own perspective and (3) it is a more complex model widely used in classification problems which is capable of solving intricate classification scenarios.

\subsection{Settings}
We performed exhaustive experiments with four different purposes:
\begin{itemize}
    \item In order to evaluate mean classification performance of the different methods.
    \item In order to evaluate generalisation capacity.
    \item In order to check how comprehensible the resulting models are.
\end{itemize}

We measured all three aspects in terms of mean values and dispersion rate. Performance and comprehensibility test were conducted following a 5-fold cross-validation scheme over four datasets (airlines, gender, IMDB and TASS). In performance-related tests, we included a stochastic method (Monte Carlo FS) that is widely used and can be helpful as a comparison between filters and wrappers. We did not include it in further scenarios since it is not in the same category than the rest.

For the generalisation test, we used a 100-fold cross-validation reversed approach. That means that each partition will have approximately $16000$ instances and we will use one partition to train and the 99 remaining ones to validate the model. With this decision, we simulated a situation in which the model has been trained with a small dataset but has to face a real word scenario. It will not be realistic to train supervised models with datasets larger than that, since expert-labelled datasets are expensive to ensemble.

All experiments where run using python 3.8.1, pandas 0.25.3, Natural Language Toolkit (NLTK) 3.4.5 and scikit-learn (sklearn) 0.22. The set of stop words that NLTK offers for each language was enriched with several custom stop words collected by ourselves. All tweets were tokenised using NTLK's \textit{TweetTokenizer} class and stemmed using \textit{SnowballStemmer} from the same package. Tweet classes were binarised.

We used the classifiers and feature selection methods present in \textit{sklearn} and we implemented those who were not available in the package. Although Information Gain (IG) and Mutual Information (MI) can be proved equal for binary classification problems, we studied them both since they are approximated with different formulas.

With the exception of those tests that explicitly evaluate some variable against the number of features, we conducted our experiments using 9 features as much, as we established earlier (see section~\ref{sec:preliminary}).

Additionally, since state-of-the-art performance is obtained when models are not subject to interpretability, we wanted to quantify baseline performance loss when using interpretable models. In consequence, we include results for ELMo in a comparative perspective despite that deep neural networks (DNNs) are not interpretable. ELMo is a language model that offers contextualised word embeddings which results in outstanding classification performance~\cite{peters_deep_2018}. The model is pretrained on the \textit{1 Billion Word Language Model Benchmark}\footnote{http://www.statmt.org/lm-benchmark/}, a large corpus of about 30 million sentences~\cite{chelba_one_2014}. Experiments were run using Tensorflow Hub ELMo v3 model\footnote{https://tfhub.dev/google/elmo/3} with the same preprocessing steps and datasets. We added a dense layer of 256 nodes followed by an output layer of 2 neurons to encode the classes. We do not restrict the number of features used by the model because (1) the model is not interpretable anymore so it is not necessary to improve its comprehensibility and since (2) we wanted the comparison to be as realistic as possible.
\section{Results and Discussion}
\label{sec:results}
In this section, we analyse the results of the experiments described above. We evaluated classification performance, stability across experiments, generalisation capacity and comprehensibility. Generally, plots show dataset mean results meanwhile tables have a more detailed perspective (separated by datasets). Conducted experiments depend on four dimensions: dataset, feature selection method, number of features, and classifier. Although, we have set fixed conditions for some of them (see section~\ref{sec:experiments}).

\subsection{Classification Performance}
\begin{figure}
    \centering
    \includegraphics[width=\columnwidth]{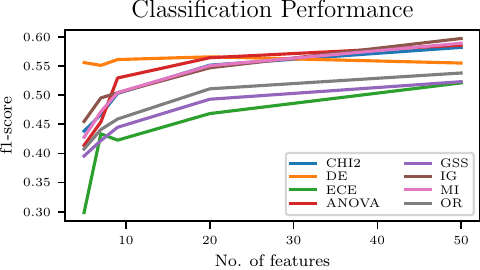}
    \caption{Classification performance measured with $f1$-score when varying the number of features. Results shown are aggregated by dataset and classifier. Our proposal has a better performance under 20 features.}
    \label{fig:f1aggregated}
\end{figure}

There are several ways of measuring how good a classification model is, but we are going to focus on $f1$-score because its ability to work with imbalanced classes. Table~\ref{tab:classificationperformance} shows accuracy, precision, recall, $f1$-score and area under the curve ROC. Presented values correspond to the mean between cross-validation folds, datasets and classifiers. Figure~\ref{fig:f1aggregated} shows $f1$-score evolution against the number of features. In order to answer the question ``\textit{what is the performance loss when using interpretable pipelines with respect to state-of-the-art approaches?}'', we include results for an ELMo-based deep neural network (DNN) that, despite that it is not interpretable, is suitable to represent baseline performance of deep contextualised language models.

Features selected by Discriminatory Expressions (DE) are more useful to train a classifier with a reduced feature space. As the number of features increases, the rest of methods get closer and even surpass DE in some cases. This happens when the number of features is higher than the amount that we can consider feasible to be managed by humans (see section~\ref{sec:preliminary}). When varying the classifier, table~\ref{tab:perclassifier} shows that our proposal works better than the rest for 3 out of 4 classifiers. DNN model does not show a significant improvement of the classification performance. It overtakes our proposal by $0.0035$ in area-under-the-curve ROC but it is behind our approach by $0.325$ in $f1$-score.

\begin{table*}
    \centering
    \caption{Classification performance of our proposal (DE) against the rest of the selected methods measured with accuracy, precision, recall, $f1$-score and area-under-curve ROC when training with 9 features. $^*$The number of features used by ELMo was not restricted.}
    \label{tab:classificationperformance}
    \footnotesize\addtolength{\tabcolsep}{-3pt}
    \begin{tabular}{lrrrrrrrrrr}
    \toprule
    {} & \multicolumn{2}{c}{accuracy} & \multicolumn{2}{c}{precision} & \multicolumn{2}{c}{recall} & \multicolumn{2}{c}{$f1$-score} & \multicolumn{2}{c}{ROC AUC} \\
    FS method &   mean &    std &   mean &    std &   mean &    std &   mean &    std &   mean &    std \\
    \midrule
    CHI2    & 0.6692 & 0.1273 & 0.6763 & 0.1461 & 0.4979 & 0.2845 & 0.5030 & 0.1472 & 0.6226 & 0.0720 \\
    DE      & 0.6869 & 0.1156 & 0.6309 & 0.0955 & 0.5367 & 0.1472 & \textbf{0.5610} & \textbf{0.0624} & 0.6370 & 0.0540 \\
    ECE     & 0.6353 & 0.1267 & 0.5872 & 0.1196 & 0.3950 & 0.2485 & 0.4228 & 0.1525 & 0.5714 & 0.0607 \\
    f-ANOVA & 0.6787 & 0.1198 & 0.6829 & 0.1370 & 0.5282 & 0.2754 & 0.5295 & 0.1306 & 0.6306 & 0.0680 \\
    GSS     & 0.6608 & 0.1311 & 0.6691 & 0.1168 & 0.3869 & 0.2189 & 0.4449 & 0.1322 & 0.6102 & 0.0743 \\
    IG      & 0.6793 & 0.1194 & 0.7006 & 0.1350 & 0.4756 & 0.2637 & 0.5040 & 0.1335 & 0.6286 & 0.0695 \\
    MI      & 0.6784 & 0.1223 & 0.7014 & 0.1456 & 0.4797 & 0.2698 & 0.5044 & 0.1379 & 0.6280 & 0.0694 \\
    OR      & 0.6500 & 0.1229 & 0.6033 & 0.1079 & 0.4214 & 0.2200 & 0.4591 & 0.1330 & 0.5927 & 0.0649 \\
    \midrule
    ELMo$^*$    & 0.6811 & 0.1480 & 0.6908 & 0.1460 & 0.5403 & 0.3074 & 0.5285 & 0.2076 & 0.6405 & 0.0995 \\

    \bottomrule
    \end{tabular}
\end{table*}

\begin{table*}
    \centering
    \caption{Detailed performance of the feature selection methods for each classifier, trained with 9 features. Listed metrics are the mean values across datasets and folds. Our proposal has the best results for 3 out of 4 classifiers.}
    \label{tab:perclassifier}
    \footnotesize\addtolength{\tabcolsep}{-3pt}
    \begin{tabular}{lrrrrrrrrrr}
    \toprule
            & \multicolumn{2}{c}{accuracy} & \multicolumn{2}{c}{precision} & \multicolumn{2}{c}{recall} & \multicolumn{2}{c}{$f1$-score} & \multicolumn{2}{c}{ROC AUC} \\
            &   mean &    std &   mean &    std &   mean &    std &   mean &    std &   mean &    std \\
    Classifier &        &        &        &        &        &        &        &        &        &        \\
    \midrule
    \textbf{CHI2}\\
    \hspace{2mm}DT & 0.6805 & 0.1189 & 0.6912 & 0.1333 & 0.5277 & 0.2667 & 0.5346 & 0.1128 & 0.6329 & 0.0679 \\
    \hspace{2mm}LR & 0.6782 & 0.1221 & 0.6951 & 0.1382 & 0.4566 & 0.2757 & 0.4881 & 0.1548 & 0.6257 & 0.0681 \\
    \hspace{2mm}RF & 0.6806 & 0.1189 & 0.6906 & 0.1328 & 0.5289 & 0.2658 & 0.5355 & 0.1127 & 0.6336 & 0.0688 \\
    \hspace{2mm}kNN & 0.6374 & 0.1492 & 0.6283 & 0.1743 & 0.4782 & 0.3346 & 0.4539 & 0.1874 & 0.5981 & 0.0810 \\
    \textbf{DE}\\
    \hspace{2mm}DT & 0.6791 & 0.0991 & 0.5970 & 0.1032 & 0.5184 & 0.1221 & 0.5401 & \textbf{0.0560} & 0.6359 & 0.0449 \\
    \hspace{2mm}LR & 0.7070 & 0.1209 & 0.6871 & 0.0930 & 0.5342 & 0.1157 & \textbf{0.5854} & \textbf{0.0346} & 0.6571 & 0.0561 \\
    \hspace{2mm}RF & 0.6970 & 0.1181 & 0.6557 & 0.0832 & 0.5302 & 0.1167 & \textbf{0.5721} & \textbf{0.0387} & 0.6501 & 0.0559 \\
    \hspace{2mm}kNN & 0.6644 & 0.1267 & 0.5837 & 0.0646 & 0.5641 & 0.2167 & \textbf{0.5464} & \textbf{0.0946} & 0.6049 & 0.0466 \\
    \textbf{ECE}\\
    \hspace{2mm}DT & 0.6398 & 0.1259 & 0.5942 & 0.1281 & 0.3978 & 0.2420 & 0.4340 & 0.1351 & 0.5769 & 0.0610 \\
    \hspace{2mm}LR & 0.6506 & 0.1201 & 0.6273 & 0.0948 & 0.3800 & 0.2730 & 0.4085 & 0.1913 & 0.5802 & 0.0577 \\
    \hspace{2mm}RF & 0.6413 & 0.1252 & 0.5909 & 0.1205 & 0.4222 & 0.2491 & 0.4487 & 0.1381 & 0.5781 & 0.0604 \\
    \hspace{2mm}kNN & 0.6095 & 0.1389 & 0.5365 & 0.1209 & 0.3799 & 0.2415 & 0.3999 & 0.1430 & 0.5503 & 0.0624 \\
    \textbf{f-ANOVA}\\
    \hspace{2mm}DT & 0.6822 & 0.1181 & 0.6797 & 0.1360 & 0.5748 & 0.2709 & \textbf{0.5587} & 0.1061 & 0.6354 & 0.0656 \\
    \hspace{2mm}LR & 0.6830 & 0.1169 & 0.6842 & 0.1318 & 0.5476 & 0.2693 & 0.5468 & 0.1073 & 0.6325 & 0.0607 \\
    \hspace{2mm}RF & 0.6826 & 0.1177 & 0.6796 & 0.1353 & 0.5756 & 0.2711 & 0.5593 & 0.1064 & 0.6360 & 0.0656 \\
    \hspace{2mm}kNN & 0.6670 & 0.1324 & 0.6883 & 0.1521 & 0.4149 & 0.2741 & 0.4530 & 0.1674 & 0.6183 & 0.0810 \\
    \textbf{GSS}\\
    \hspace{2mm}DT & 0.6792 & 0.1178 & 0.6973 & 0.0922 & 0.3321 & 0.1435 & 0.4311 & 0.1344 & 0.6251 & 0.0738 \\
    \hspace{2mm}LR & 0.6814 & 0.1163 & 0.7075 & 0.0843 & 0.3104 & 0.1045 & 0.4235 & 0.1158 & 0.6117 & 0.0521 \\
    \hspace{2mm}RF & 0.6806 & 0.1173 & 0.6928 & 0.0872 & 0.3407 & 0.1461 & 0.4379 & 0.1372 & 0.6269 & 0.0742 \\
    \hspace{2mm}kNN & 0.6020 & 0.1580 & 0.5789 & 0.1472 & 0.5644 & 0.3172 & 0.4872 & 0.1384 & 0.5769 & 0.0863 \\
    \textbf{IG}\\
    \hspace{2mm}DT & 0.6818 & 0.1196 & 0.6952 & 0.1347 & 0.5307 & 0.2628 & 0.5394 & 0.1059 & 0.6346 & 0.0696 \\
    \hspace{2mm}LR & 0.6774 & 0.1246 & 0.6971 & 0.1302 & 0.3964 & 0.2510 & 0.4548 & 0.1651 & 0.6230 & 0.0751 \\
    \hspace{2mm}RF & 0.6819 & 0.1197 & 0.6948 & 0.1344 & 0.5326 & 0.2628 & 0.5405 & 0.1065 & 0.6355 & 0.0710 \\
    \hspace{2mm}kNN & 0.6762 & 0.1209 & 0.7155 & 0.1475 & 0.4426 & 0.2673 & 0.4813 & 0.1338 & 0.6213 & 0.0650 \\
    \textbf{MI}\\
    \hspace{2mm}DT & 0.6845 & 0.1185 & 0.7156 & 0.1407 & 0.4987 & 0.2546 & 0.5253 & 0.1104 & 0.6365 & 0.0677 \\
    \hspace{2mm}LR & 0.6813 & 0.1227 & 0.7166 & 0.1421 & 0.3965 & 0.2423 & 0.4601 & 0.1573 & 0.6275 & 0.0712 \\
    \hspace{2mm}RF & 0.6847 & 0.1185 & 0.7153 & 0.1403 & 0.5001 & 0.2547 & 0.5262 & 0.1111 & 0.6373 & 0.0688 \\
    \hspace{2mm}kNN & 0.6632 & 0.1352 & 0.6582 & 0.1585 & 0.5236 & 0.3191 & 0.5060 & 0.1618 & 0.6106 & 0.0706 \\
    \textbf{OR}\\
    \hspace{2mm}DT & 0.6543 & 0.1245 & 0.6082 & 0.1058 & 0.4565 & 0.2148 & 0.4897 & 0.1058 & 0.6014 & 0.0655 \\
    \hspace{2mm}LR & 0.6576 & 0.1237 & 0.6288 & 0.0949 & 0.3574 & 0.2227 & 0.4161 & 0.1694 & 0.5921 & 0.0672 \\
    \hspace{2mm}RF & 0.6564 & 0.1232 & 0.6098 & 0.1021 & 0.4628 & 0.2133 & 0.4950 & 0.1034 & 0.6039 & 0.0648 \\
    \hspace{2mm}kNN & 0.6317 & 0.1259 & 0.5665 & 0.1235 & 0.4087 & 0.2257 & 0.4355 & 0.1328 & 0.5734 & 0.0615 \\
    \bottomrule
    \end{tabular}
\end{table*}

\begin{figure}
    \centering
    \includegraphics[width=\columnwidth]{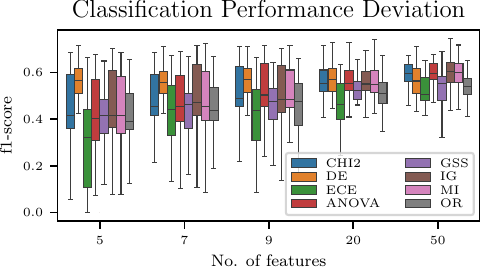}
    \caption{$f1$-score dispersion between iterations for each fold, classifier and dataset while varying the number of features. Our proposal shows a lower dispersion and, generally, better performance while working with a low number of features.}
    \label{fig:dispersion_f1}
\end{figure}

In order to prove that a new feature selection method performs well, it is also necessary to consider measures of dispersion. In the aforementioned tables, it is possible to check that DE dispersion is the lowest among the rest. Table~\ref{tab:dispersion_additional} shows additional deviation measurements. Discriminatory Expressions (DE) is significantly more stable than the rest of them. Figure~\ref{fig:dispersion_f1} depicts the dispersion of the performance for each feature selection method while varying the number of features. They show that our proposal is more stable through different classifiers, datasets and folds, especially while working with a low number of features. DNN model variability is also higher than the rest of them, just behind Expected Cross Entropy based models, which are the most disperse.

Discriminatory Expressions outperforms the rest of the selected filters when varying the dataset and/or the classifier, while requiring a low number of features. It is also the most stable method. This makes it a neat approach in order to achieve good classification performance in a context where actual interpretability is required.

\begin{table}
    \centering
    \caption{Further measurements of performance dispersion for our proposal (DE) against the rest of the selected methods measured with range, interquartile range (IQR), standard deviation (STD) and coefficient of variation (CV) when training with 9 features. $^*$The number of features used by ELMo was not restricted.}
    \label{tab:dispersion_additional}
    \begin{tabular}{lrrrr}
    \toprule
    {} &  range &    IQR &    STD &  CV (\%) \\
    FS method      &        &        &        &         \\
    \midrule
    CHI2    & 0.6730 & 0.1732 & 0.1472 & 29.2623 \\
    DE      & \textbf{0.2977} & \textbf{0.1008} & \textbf{0.0624} & \textbf{11.1313} \\
    ECE     & 0.5772 & 0.2161 & 0.1525 & 36.0694 \\
    f-ANOVA & 0.6276 & 0.1829 & 0.1306 & 24.6674 \\
    GSS     & 0.4964 & 0.1330 & 0.1322 & 29.7014 \\
    IG      & 0.5860 & 0.2020 & 0.1335 & 26.4790 \\
    MI      & 0.6420 & 0.1625 & 0.1379 & 27.3400 \\
    OR      & 0.5174 & 0.1801 & 0.1330 & 28.9810 \\
    \midrule
    ELMo$^*$    & 0.7904 & 0.2368 & 0.2076 & 39.2822 \\
    \bottomrule
    \end{tabular}
\end{table}

\subsection{Generalisation Capacity}
In order to measure how good a method models reality when training with small datasets, we used a reversed 100-fold cross-validation scheme. We split our data into 100 partitions and we measure the performance when classifying 99 partitions having trained with only one of them. This means that, for each fold, the model trained with approximately 16k instances and it was asked then to classify 1.5 million instances. As in the rest of the experiments, we considered 9 features.
We considered mean and dispersion rates for accuracy, precision, recall, $f1$-score and area under the curve ROC. Table~\ref{tab:generalisation} shows the behaviour of the feature selection method for our biggest dataset (Sentiment140). Results show that our proposal not only has the best mean generalisation capacity but also the lowest deviation.

\begin{table*}
    \centering
    \caption{Mean generalisation capacity and its deviation for a reversed 100-fcv using 9 features with Sentiment140 dataset. Results are the mean between folds and classifiers. Our proposal has a slightly superior $f1$-score than the rest. It is also the most stable one.}
    \label{tab:generalisation}
    \footnotesize\addtolength{\tabcolsep}{-3pt}
    \begin{tabular}{lrrrrrrrrrr}
    \toprule
    {} & \multicolumn{2}{c}{accuracy} & \multicolumn{2}{c}{precision} & \multicolumn{2}{c}{recall} & \multicolumn{2}{c}{$f1$-score} & \multicolumn{2}{c}{ROC AUC} \\
    FS method &   mean &    std &   mean &    std &   mean &    std &   mean &    std &   mean &    std \\
    \midrule
    CHI2    & 0.5781 & 0.0071 & 0.5672 & 0.0672 & 0.8805 & 0.2320 & 0.6553 & 0.1218 & 0.5781 & 0.0071 \\
    DE      & 0.6526 & 0.0161 & 0.6313 & 0.0198 & 0.7446 & 0.0988 & \textbf{0.6784} & \textbf{0.0456} & 0.6526 & 0.0161 \\
    ECE     & 0.5370 & 0.0068 & 0.5790 & 0.0553 & 0.4557 & 0.2946 & 0.4437 & 0.1674 & 0.5371 & 0.0068 \\
    f-ANOVA & 0.5776 & 0.0097 & 0.5684 & 0.0702 & 0.8750 & 0.2378 & 0.6517 & 0.1312 & 0.5776 & 0.0097 \\
    GSS     & 0.5865 & 0.0208 & 0.7009 & 0.0438 & 0.3324 & 0.1274 & 0.4367 & 0.0509 & 0.5865 & 0.0208 \\
    IG      & 0.5784 & 0.0085 & 0.5605 & 0.0606 & 0.9092 & 0.1936 & 0.6682 & 0.1092 & 0.5784 & 0.0085 \\
    MI      & 0.5784 & 0.0086 & 0.5650 & 0.0670 & 0.8941 & 0.2176 & 0.6607 & 0.1203 & 0.5784 & 0.0086 \\
    OR      & 0.5549 & 0.0155 & 0.6559 & 0.0645 & 0.3059 & 0.1941 & 0.3819 & 0.0944 & 0.5549 & 0.0155 \\
    \bottomrule
    \end{tabular}
\end{table*}

\subsection{Comprehensibility}
As we established earlier, there is no global measure of interpretability. In this section, we will perform an analysis on the practical interpretability of the models that result from each feature selection method.

$k$-nearest neighbours (kNN) get the $k$ most similar neighbours to the one that is being classified and outputs the majority class. The interpretation is straightforward, since it is a mere comparison. Hence, the fewer neighbours you need to consider, the better. Likewise, the lower the number of features the better, because that means that there will be less dimensions when comparing similarities. In order to measure overall performance versus complexity, we will use equation~\ref{eq:knncomprehensibility}:
\begin{equation}
    \label{eq:knncomprehensibility}
    \text{comprehensibility rate} = \sum_k{\frac{f1\text{-score}}{k}}    
\end{equation}
since the comprehensibility decays with the number of neighbours $k$ considered.

\begin{figure}[tbp]
    \centering
    \includegraphics[width=\columnwidth]{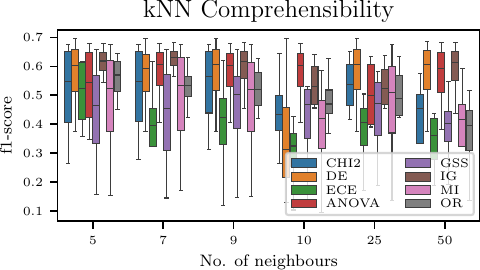}
    \caption{Performance comparison between kNN models generated using different feature sets. The plot shows the number of neighbours considered (less is better) against $f1$-score. Our proposal performance remain between the top approaches for this classifier while needing a low number of features and neighbours.}
    \label{fig:knncomprehensibility}
\end{figure}

Table~\ref{tab:knncomplexity} shows performance results for different number of neighbours and $9$ features. Figure~\ref{fig:knncomprehensibility} shows different dispersion measures between folds and datasets for $f1$-score values with respect to the number of considered neighbours. Discriminatory Expressions (DE) method has its better comprehensibility rate grouped under $9$ neighbours, without significant difference than when using $5$ of them.

\begin{table*}
    \centering
    \caption{Classification performance measure with $f1$-score against $k$-nearest neighbours (kNN) feature selection method. The lower the number of neighbours and the higher the $f1$-score, the better.}
    \label{tab:knncomplexity}
    \footnotesize\addtolength{\tabcolsep}{-3pt}
    \begin{tabular}{lrrrrrrrrrr|r}
    \toprule
    k & \multicolumn{2}{c}{5} & \multicolumn{2}{c}{7} & \multicolumn{2}{c}{9} & \multicolumn{2}{c}{10} & \multicolumn{2}{c|}{25} & $\sum\frac{f1}{k}$\\
    FS method &   mean &    std &   mean &    std &   mean &    std &   mean &    std &   mean &    std & mean \\
    \midrule
    CHI2    & 0.5091 & 0.19 & 0.5120 & 0.18 & 0.5285 & 0.16 & 0.4442 & 0.15 & 0.5353 & 0.10 & 0.2995 \\
    DE      & 0.5693 & 0.14 & 0.5641 & 0.13 & 0.5702 & 0.14 & 0.3616 & 0.24 & 0.5722 & 0.14 & 0.3169 \\
    ECE     & 0.5056 & 0.12 & 0.3808 & 0.20 & 0.3957 & 0.20 & 0.2946 & 0.13 & 0.3721 & 0.14 & 0.2438 \\
    f-ANOVA & 0.5285 & 0.15 & 0.5749 & 0.12 & 0.5729 & 0.11 & 0.5722 & 0.11 & 0.5058 & 0.12 & 0.3289 \\
    GSS     & 0.4360 & 0.21 & 0.4278 & 0.22 & 0.4513 & 0.21 & 0.3987 & 0.18 & 0.4295 & 0.17 & 0.2555 \\
    IG      & 0.6148 & 0.05 & 0.6271 & 0.04 & 0.5935 & 0.09 & 0.5391 & 0.08 & 0.5351 & 0.08 &\textbf{ 0.3538} \\
    MI      & 0.4691 & 0.23 & 0.4781 & 0.22 & 0.4676 & 0.23 & 0.3799 & 0.20 & 0.4583 & 0.23 & 0.2704 \\
    OR      & 0.5595 & 0.08 & 0.5293 & 0.08 & 0.5228 & 0.09 & 0.4838 & 0.10 & 0.5094 & 0.10 & 0.3144 \\
    \bottomrule
    \end{tabular}
\end{table*}

Decision Trees (DT) interpretation relies not only in the number of features selected but also in (1) the number of leaves (equivalent to number of rules) and (2) the length of the path from the root to each leaf node. In both cases, less is better. We consider that the desired decision tree will be have approximately 10 rules of 5 clauses each. Hence, we will be measuring the complexity with equation~\ref{eq:dtcomplexity}.

\begin{equation}
    \label{eq:dtcomplexity}
    \text{complexity} = \frac{\text{no. rules}}{\text{baseline rules}}\left(\frac{\text{no. clauses}}{\text{baseline clauses}}\right)^2
\end{equation}

We used the squared relation between the number of clauses and the desired one because the length of the rule would affect more to the comprehensibility than the number of rules. The relation between accuracy and complexity will respond to equation~\ref{eq:dtcomprehensibility}:

\begin{equation}
    \label{eq:dtcomprehensibility}
    \text{comprehensibility rate} = 100\cdot\frac{f1\text{-score}}{complexity}
\end{equation}

Table~\ref{tab:dtcomplexity} shows $f1$-score against complexity measures (number of leaves and mean length of the path). Our proposal is the only filter that manage to score good in classification performance while keeping a simple DT, both in terms of mean and dispersion measures.

\begin{figure}[tbp]
    \centering
    \includegraphics[width=\columnwidth]{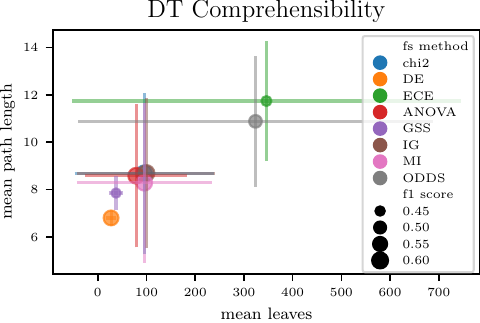}
    \caption{Complexity measures comparison between DT models generated using different feature sets. The plot shows the mean number of leaves (fewer is better) as well as the mean length of the paths to those leaves (less is better). The size of each marker stands for the $f1$-score and the error bars represent the dispersion (standard deviation) for each dimension. Our proposal keeps the lowest complexity of the model while keeping one of the best performances.}
    \label{fig:dtcomprehensibility}
\end{figure}

Figure~\ref{fig:dtcomprehensibility} shows centrality and dispersion measurements on the number of rules and their length. It also shows the classification performance. It is possible to notice that Discriminatory Expressions method is (1) the feature set that produces the simpler trees (the closer to the bottom left corner, the better), (2) the most stable in both dimensions (shortest whiskers) and (3) the feature set that produces one of the best performing models (dot size).

\begin{table*}[tbp]
    \centering
    \caption{Performance and complexity measures of DT-based models. Number of leaves (rules) and path length (clauses) can be used to compare how comprehensible are the models that result from training with different feature sets.}
    \label{tab:dtcomplexity}
    \footnotesize\addtolength{\tabcolsep}{-3pt}
    \begin{tabular}{lrrrrrr|rr}
    \toprule
    {} & \multicolumn{2}{c}{$f1$-score} & \multicolumn{2}{c}{rules} & \multicolumn{2}{c|}{clauses} & complexity & compr. rate \\
    FS method &   mean &    std &     mean &      std &    mean &    std &       mean &              mean \\
    \midrule
    CHI2    & 0.5995 & 0.08 &  95.2500 & 142.71 &  8.6858 & 3.41 &    28.74 &            2.08 \\
    DE      & 0.5711 & 0.14 &  27.2500 &  10.43 &  6.8067 & 0.32 &     \textbf{5.05} &           \textbf{11.30} \\
    ECE     & 0.4623 & 0.13 & 346.2500 & 398.57 & 11.7405 & 2.53 &   190.90 &            0.24 \\
    f-ANOVA & 0.5939 & 0.09 &  78.7500 & 104.26 &  8.5890 & 3.02 &    23.23 &            2.55 \\
    GSS     & 0.4486 & 0.19 &  37.5000 &  14.88 &  7.8571 & 0.70 &     9.26 &            4.84 \\
    IG      & 0.6051 & 0.08 &  99.7500 & 141.43 &  8.6970 & 3.15 &    30.17 &            2.00 \\
    MI      & 0.5927 & 0.09 &  95.5000 & 138.48 &  8.2910 & 3.37 &    26.25 &            2.25 \\
    OR      & 0.5167 & 0.12 & 323.7500 & 363.75 & 10.8829 & 2.75 &   153.37 &            0.33 \\
    \bottomrule
    \end{tabular}
\end{table*}

Finally, Random Forests (RF) are basically collections of DTs. Their complexity can be evaluated in the same way than decision trees, but it is necessary to consider an additional dimension, that is the number of trees in the forest (fewer is better). Hence, complexity measure will now reflect this new dimension as in equation~\ref{eq:rfcomplexity}:
\begin{multline}
    \label{eq:rfcomplexity}
    \text{complexity} = \\\frac{\text{no. trees}\cdot\text{no. rules}}{\text{baseline trees}\cdot\text{baseline rules}}\left(\frac{\text{no. clauses}}{\text{baseline clauses}}\right)^2
\end{multline}

\begin{figure}[tbp]
    \centering
    \includegraphics[width=\columnwidth]{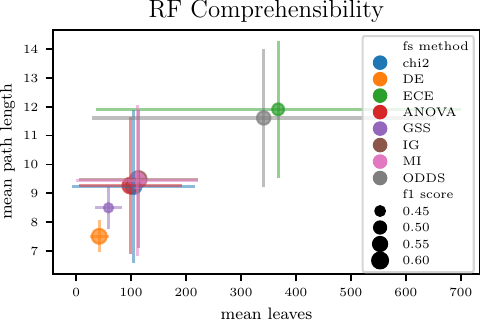}
    \caption{Comprehensibility comparison between the inner trees of RF models generated using different feature sets. The plot shows the mean number of leaves (less is better) as well as the mean length of the paths to those leaves (less is better). The size of each marker stands for the $f1$-score and the error bars represent the dispersion (standard deviation) for each dimension. Our proposal keeps the lowest complexity of the model while keeping one of the best performances and stabilities.}
    \label{fig:rfcomprehensibility}
\end{figure}

\begin{figure}[tbp]
    \centering
    \includegraphics[width=\columnwidth]{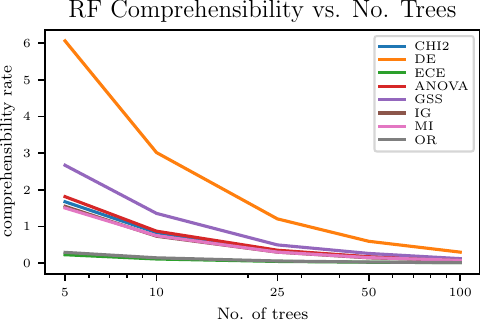}
    \caption{Comprehensibility rate for RF models when varying the number of trees in each forest. The rate decreases as the number of trees increases.}
    \label{fig:rfrate}
\end{figure}

The comprehensibility rate will not suffer further changes. Table~\ref{tab:rfcomplexity} shows $f1$-score and complexity measures. Discriminatory Expressions method manages to perform well in terms of accuracy while keeping a minimum deviation. Figure~\ref{fig:rfcomprehensibility} shows centrality and dispersion measurements for each feature set, while figure~\ref{fig:rfrate} shows the rate between accuracy and complexity. Our proposal has the best behaviour among the tested methods.

\begin{table*}
    \centering
    \caption{Performance and complexity measures of RF-based models when training with 5 trees. Number of leaves (rules) and path length (clauses) can be used to compare how comprehensible are the models that result from training with different features selection mechanisms.}
    \label{tab:rfcomplexity}
    \footnotesize\addtolength{\tabcolsep}{-3pt}
    \begin{tabular}{lrrrrrr|rr}
    \toprule
    {} & \multicolumn{2}{c}{$f1$-score} & \multicolumn{2}{c}{leaves} & \multicolumn{2}{c|}{length} & complexity & compr. rate \\
    FS method &   mean &    std &     mean &      std &    mean &    std &       mean &              mean \\
    \midrule
    CHI2    & 0.6039 & 0.08 & 105.3000 & 126.73 &  9.2564 & 3.18 &    36.08 &            1.67 \\
    DE      & 0.5707 & 0.14 &  41.9000 &  19.93 &  7.4953 & 0.62 &     \textbf{9.41} &            \textbf{6.06} \\
    ECE     & 0.4681 & 0.13 & 368.9000 & 377.17 & 11.7672 & 2.82 &   204.32 &            0.22 \\
    F-ANOVA & 0.5954 & 0.10 &  98.3000 & 104.34 &  9.1386 & 2.47 &    32.83 &            1.81 \\
    GSS     & 0.4485 & 0.18 &  58.6500 &  28.79 &  8.4637 & 0.82 &    16.80 &            2.66 \\
    IG      & 0.6102 & 0.08 & 111.2000 & 118.40 &  9.4097 & 2.30 &    39.38 &            1.54 \\
    MI      & 0.5937 & 0.09 & 109.8500 & 124.97 &  9.4794 & 3.01 &    39.48 &            1.50 \\
    OR      & 0.5309 & 0.13 & 341.1000 & 349.87 & 11.5126 & 2.68 &   180.83 &            0.29 \\
    \bottomrule
    \end{tabular}
\end{table*}

\section{Contributions and Limitations}
\label{sec:limitations}
In this section, we summarise the contributions and limitations of our work.

We proposed the use of expression as features instead of words or other tokens. We based this decision on the work of linguists when analysing text. They consist on sequences of words in a certain order that may have others words in between. Not only these features are more interpretable to humans but they also add certain context to document encoding that may help classifiers in disambiguation. Moreover, we take advantage of stop words, since we do not discard them immediately as most models do. Instead, we use them to specialise expressions so that they are more accurate. Most of the methods we tested proved to be more sensible to the classifier (and dataset) than ours. Discriminatory Expressions (DE) would be less disrupted by external facts like context and/or classifier.

The decision of using a filtering method is also influenced by interpretability. The notion of optimising the feature subset to the classifiers is against transparency to some extent. Wrappers adapt their selection to the performance of the subsequent classifier, while filters are based on \textit{a priori} information.

Moreover, current interpretable machine learning (ML) models are not so interpretable in actual fact. Due to the complexity of our reality, trained models result in a vast number of rules with a large number of chained clauses that are nearly impossible to understand in practice. We presented a method that is capable of maintaining an acceptable performance while diminishing the complexity of the subsequent classifier. This is specially noticeable when using Decision Trees.

However, our method selects a relevant feature set but the classification performance does not improve much with the number of features. In other words, discriminatory expressions enhance accuracy with a low number of features but reaches a performance ceiling when augmenting the number of features. We believe this may be related to target recall and relevance and may be overcome by gradually decreasing both parameters.

Summing up, we would recommend the use of our method:
\begin{itemize}
    \item When actual interpretability is required.
    \item With any classifier, but specially with decision trees.
    \item When features need to be explainable by themselves.
    \item When fewer features are preferred.
\end{itemize}
\section{Conclusions}
\label{sec:conclusions}
Nowadays, Social Media has become one of the most important sources of information and a crucial way of communication. Microblogging sites are being used to analyse certain aspects of our reality, due to their peculiarities such as promptness and versatility. However, short-text messages present handicaps when training machine learning models, such as contractions or misspelled words. In recent years, machine learning has focused mainly on improving classification performance rather than coping with these handicaps in an interpretable manner. Thus, state-of-the-art approaches are black-box models that should not be used to take or assist decisions that may have a social impact.

Despite that we may think that there are interpretable classifiers that can be used to build interpretable models, the fact is that the resulting ones are too complex to be actually interpretable. They usually require a large number of rules with several clauses, which reduces the comprehensibility since we humans have limited capacity. It is necessary to obtain new approaches that lead us to authentic transparency in artificial intelligence. Feature selection was a good starting point, provided that it is the initial step in machine learning pipelines.

We selected several methods and classifiers that are widely used, both in scientific literature and commercial solutions, and we conducted several experiments in order to perform a preliminary study to check whether current methods are good enough to be comprehensible when combined with interpretable classifiers. Our results showed that there was still room for improvement.

We developed a feature selection algorithm for microblogging contexts where interpretability is mandatory that focused in obtaining a well-structured set of significant features in order to improve the comprehensibility of the model while offering good classification performance. We used expressions (sequences of words) that are biased towards specific classes, inspired in linguists procedures when analysing text. In order to assemble those expressions, we introduced a word ranking method that we have called \textit{CF-ICF}. This method is capable of scoring words in two dimensions: statistical relevance and discriminative power (that is, bias towards an specific class).

We conducted exhaustive experiments to compare current methods with our proposal, in terms of classification accuracy, generalisation capacity and comprehensibility of the resulting models. We compared eight different feature selection methods using five different popular datasets. We followed a cross-validation scheme to ensure the integrity of the results with four different classifiers.

Results showed that our algorithm achieves a good baseline performance (without hyperparameter tuning) while being the most stable method regardless of the context and classifier used. For each iteration and fold, our algorithm yields close results, minimising deviations in almost every case, with a big difference with respect to the rest of the studied methods. It happens the same with the generalisation capacity, since our algorithm is the best both in terms of mean value and dispersion rate. It is also noticeable that classification performance does not improve significantly with the number of features, which proves that our feature set is quite relevant from the beginning.

An ELMo-based Deep Neural Network (DNN) was also included in the comparison. DNNs are state-of-the-art approaches for Natural Language problems, and we prove the baseline loss is not significant. DNN is behind our proposal with respect to $f1$-score while it is better in terms of area-under-the-curve ROC. In both cases, DE is more consistent in terms of standard deviation. Even thought it is likely that DNN improves significantly when fine tuned with larger datasets, there is no \textit{a priori} reason to choose a non-interpretable model to deal with real microblogging (or other short text documents with limited context) datasets.

As for comprehensibility, our proposal manage to score among the best for every case. When combined with decision trees, our algorithm reduces drastically the complexity of the models without loosing performance. Complexity deviation is also one the lowest of the eight selected methods, one order of magnitude lower than the majority of them. When used to train random forests, our proposal gives the best overall results of accuracy versus complexity.

\section{Future Work}
\label{sec:futurework}
There are several lines of future work, including a more in-depth evaluation of interpretability, focusing in an application-level analysis within a specific application. We also plan to perform analytical studies on hyperparameter fine-tuning to maximise both performance and comprehensibility. This is also the first step towards an interpretable pipeline, and we plan to continue proposing novel methods and/or modifications of current ones to obtain a fully comprehensible classification pipeline.

\appendix[Complementary Tables]



\ifCLASSOPTIONcompsoc
  \section*{Acknowledgments}
\else
  \section*{Acknowledgment}
\fi

This work has been financially supported by the Spanish Ministry of Economy and Competitiveness (MINECO), project FFI2016-79748-R, and cofinanced by the European Social Fund (ESF). Manuel Francisco Aparicio was supported by the FPI 2017 predoctoral programme, from the Spanish Ministry of Economy and Competitiveness (MINECO), grant number BES-2017-081202.

\ifCLASSOPTIONcaptionsoff
  \newpage
\fi



%



\bibliographystyle{IEEEtran}  
\bibliography{bibtex/bib/references.bib} 

%

\begin{IEEEbiography}[{\includegraphics[width=1in,height=1.25in,clip,keepaspectratio]{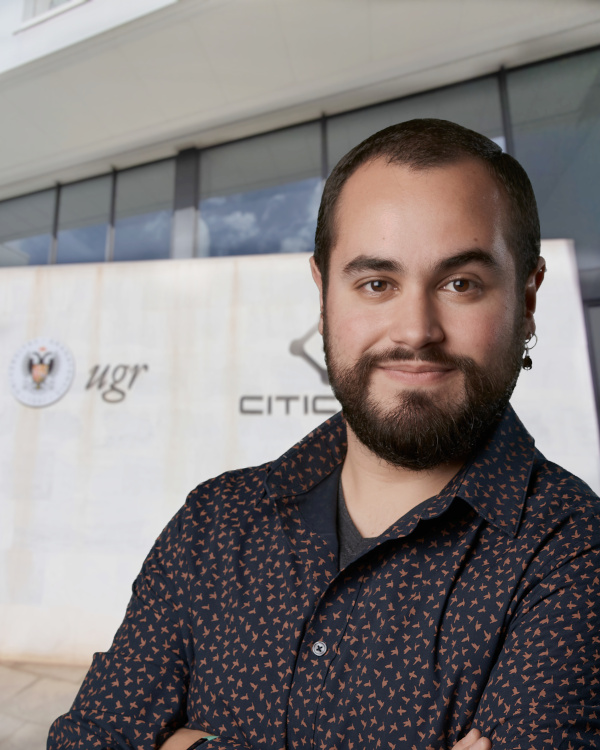}}]{Manuel Francisco} is a PhD. student and lecturer at University of Granada. In 2018, he received his MSc. in Data Science and Computer Engineering from the University of Granada, Spain. His main research field is Natural Language Processing in Social Networking Sites, focusing on interpretable techniques. He is also interested in Machine Learning, Parallel and Distributed Computing, Biologically Inspired Algorithms, Educational Innovation and literally anything that can pique his curiosity.
\end{IEEEbiography}

\begin{IEEEbiography}[{\includegraphics[width=1in,height=1.25in,clip,keepaspectratio]{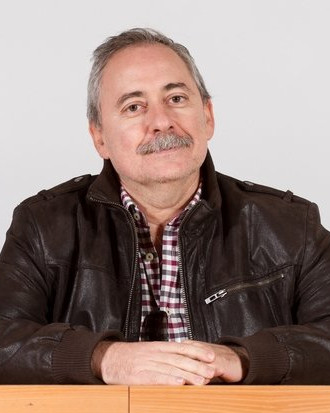}}]{Juan Luis Castro} holds the degrees of Bachelor of Mathematics (1988) and Doctor of Mathematics (1991) from the University of Granada. He is Full Professor of Computer Science and Artificial Intelligence at the University of Granada and he currently teaches at the School of Computer Engineering and Telecommunications Engineering of the University of Granada. His research interest include neural networks, fuzzy systems, machine learning, and related applications. He has published three books, more than 150 research papers in journals, conferences and books of his field of interest, and he serves as a Reviewer for some journals and international conferences on these topics. He has been Principal Investigator in several regional, national and international research projects subsidised with public funds and and several technological transfer projects directly contracted with the industry. He has been the founder of Virtual Solutions \& Artificial Intelligence, a technology-based company that was created with the goal of providing Intelligent Systems with Natural Language. He is currently the head of the research group on “Computational Intelligence” in the University of Granada.
\end{IEEEbiography}






\end{document}